\documentclass[]{JFM-FLM_Au}
\usepackage{sidecap}
\usepackage{bm}
\usepackage{soul}
\usepackage{subfig}
\newcommand{\td}{\text{d}}

\lefttitle{Gaurav \& Sharma}
\righttitle{Journal of Fluid Mechanics}

\title{Axisymmetric landslides on small planetary bodies}

\author{Kumar Gaurav \aff{1}\and Ishan Sharma \aff{1,2}}
\affiliation{\aff{1}Department of Mechanical Engineering, Indian Institute of Technology Kanpur, India
\aff{2}Space, Planetary $\&$ Astronomical Sciences and Engineering,
Indian Institute of Technology Kanpur, India}

\corresau{Kumar Gaurav, kugaurav@iitk.ac.in}

\begin{document}
\maketitle

\begin{abstract}
 We aim to understand how landslides affect the shape and rotational motion of small rubble planetary bodies. We limit ourselves to axisymmetric global landslides, and take the primordial shape of the body to also be axisymmetric. The landslides  are modeled through depth-averaging, while also incorporating the effect of the body's rotation, topographical changes due to multiple landslides, the body's non-uniform gravity field and possible surface mass shedding. The body's rotational dynamics is coupled to its shape change due to landsliding, and also includes the action of radiation torque. We  utilize our model to investigate regolith motion on actual asteroids. We then study the evolution of the shape and spin state of an initially spherical rubble asteroid due to impact-induced global landsliding events over its lifetime. We find that rotational fission is suppressed by regolith redistribution to the body's equator by landsliding. Furthermore, top shapes are rapidly formed and this may explain the abundance of top-shaped asteroids in near-Earth orbits.
\end{abstract}


%{\bf MSC Codes }  {\it(Optional)} Please enter your MSC Codes here

\section{Introduction}
\label{sec:headings}

\begin{figure}[t]
\centering  
\includegraphics[width=0.80\textwidth]{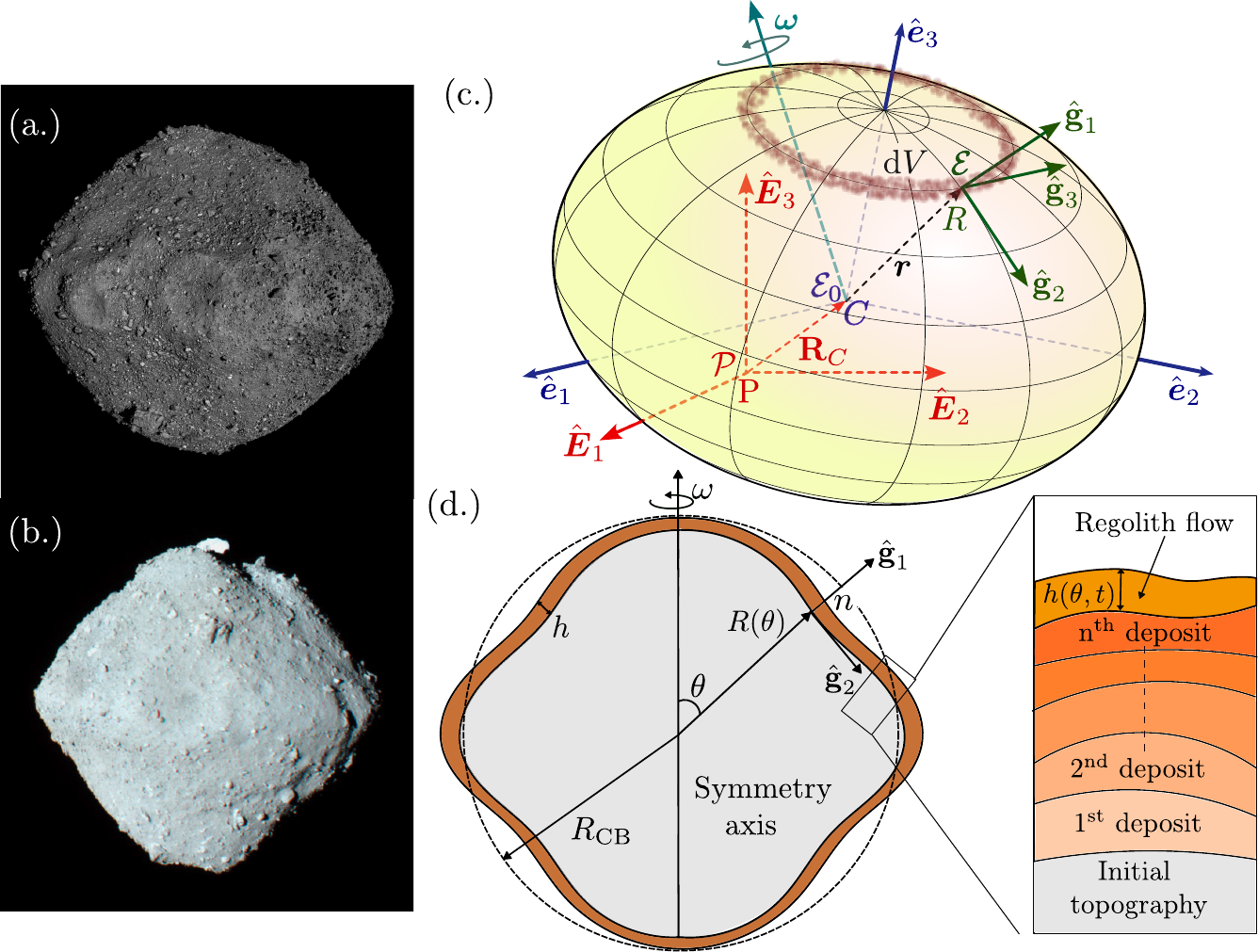}
     \caption{ Top-shaped NEAs (a) 101955 Bennu  and (b) 162173 Ryugu. The  granular surface is striking with visible rocks and large boulders. Source: NASA and JAXA. (c) Schematic of a rotating planetary body with angular velocity $\bm{\omega}$. Three different coordinate systems are indicated and these are discussed in the main text.  The differential volume element $\td V$ of an axisymmetric is shown. (d) Landslide (regolith flow) on an initially spherical body after multiple axisymmetric landsliding events.  }
     \label{fig:schematic}
\end{figure}

Small Solar System bodies, such as sub-kilometre sized near-Earth asteroids (NEAs), are observed to have grain rich -- ``sandy" -- surfaces, populated mostly by very small grains (fines) and rock, along with several large boulders; see, e.g. the surface of asteroids Bennu and Ryugu in Fig.~\ref{fig:schematic}. This observation may be related to the fact that many of these objects are believed to be ``rubble piles", i.e. they are granular aggregates that are held together primarily by self gravity (Walsh 2018)\nocite{walsh2018}. Because such small bodies lack an atmosphere, all impactors, irrespective of their size, are able to reach their surface. Consequently, these bodies experience a large number of impacts during their lifetime and, given their small size, even modest impacts may initiate seismic events that are large enough to cause global landslides. Landslides  redistribute regolith -- surface grains -- which both reshapes and textures these bodies: including crater erosion, grain sorting, and formation of ridges  and terraces. Crucially, shape changes  modify the body's mass moment of inertia and, so,  its rotation state in order to conserve the system's angular momentum. Thus, landsliding  events over the lifetime of a small Solar System body contribute greatly to changing its shape and rotation state. 

Generally, regolith motion on small planetary bodies is investigated through discrete element (DE) simulations \citep{cundall1979discrete}, and an example is the recent work of \cite{song2024integrated}. Such simulations are computationally expensive and, so, are not best suited to study shape and spin evolution of rubble piles over their lifetime, which will involve resolving many landslide events. This motivates the present work where we model landslides on small planetary bodies by suitably extending continuum models for shallow terrestrial landslides on arbitrary topography \citep{savage1989motion, gray1999gravity, cms/1109868726, pudasaini2007avalanche, Luca2016}. We limit ourselves to global axisymmetric landslides on an initially axisymmetric shape, but allow the topography to be otherwise irregular across latitudes. We incorporate the effects of the body's rotation and non-central gravitational field, as well as couple the shape changes and consequent mass redistribution to the body's rotational dynamics.

 Previous efforts in this direction include the study of two-dimensional landslides on a rotating ellipse by  \cite{gaurav2021granular} and axisymmetric landslides on a top-shaped body by \cite{banik_top}.  Here, we extend this modeling effort significantly to incorporate global axisymmetric landsliding on a body whose initial shape was {\em any} surface of revolution, while also updating the topography after every landslide. This greatly expands the applicability to realistic bodies, as well as paves the way for the fully general case of local landslides on an arbitrary body.
 
 % Even though, the axisymmetric case of landslide and shape of asteroids is unrealistic, it leads to simple equation and hence gives great physical insights. It is able to capture many physics and can easily be extended for general case, for example using the system of equations of Iona and Francois. 
 
 The paper is organized as follows. In Sec. 2, we  provide general set of equations  that solve for the regolith flow during a landslide and also couple it to the body's rotational dynamics. These are simplified for an axisymmetric system in Sec. 3. We provide illustrative results and applications to real asteroids in Sec. 4, before concluding.

\vspace{-0.5cm}
\section{General governing equations}\label{sec:landslides}

We  extend the approach of \cite{banik_top} to model regolith motion on small planetary bodies. The body itself is taken to comprise of grains that may be broadly distinguished into two regions: a solid central body (CB) where grains are consolidated and/or compacted and/or sintered, and a mobile layer of regolith that arises when landslides occur. Thus, in between two landsliding events the entire body is the CB. The shape and size of the CB is modified when regolith is mobilized, flows and stops. The rotational dynamics of the system will incorporate effects of the system's changing mass moment of inertia. For completeness, we will first present the  complete set of equations governing the coupled motion of the CB and the flowing regolith. Subsequently, we will introduce  assumptions that will simplify the presentation,  but will still be useful for realistic  applications.

%When landslides occur the body consists of a solid CB and a flowing regolith layer. Thus, we cannot strictly ascribe to the body an angular velocity, at least in the sense of a rigid body. However, landsliding events are intermittent, with each landslide remaining active for a duration of time much shorter than the rotational dynamics timescale and, moreover, landslides are typically of depth much smaller in comparison to the body's size. Thus, we may identify the CB's rotational motion with that of the body itself.

Figure~\ref{fig:schematic}c shows the body rotating with an angular velocity $\bm{\omega}$ that may or may not be aligned with a principal axis of the moment of inertia tensor.  There are three coordinate systems (CS) shown in Fig.~\ref{fig:schematic}c, viz. the inertial CS $\{ \mathcal{P},P,\hat{\bm{E}}_i\}$ with origin at some space-fixed point $P$ and unit vectors $\hat{\bm E}_i$; the body-fixed CS (BFCS)   $\{\mathcal{E}_0,C,\hat{\bm{e}}_i\}$, which is attached to the CB and is defined by its principal axes $\hat{\bm{e}}_i$ and $C$ lies at the CB's center of mass (CoM); and $\{ \mathcal{E},\, R,\, \hat{\bm{g}}_i\}$ is a curvilinear CS  defined by the natural unit basis $\hat{\bm g}_i$ associated with the body's surface. Henceforth, we  represent all vector quantities measured in the inertial frame $\mathcal{P}$ and the BFCS by, respectively, bold upper and lower case letters. 

We take the flowing regolith to be an incompressible continuum, for which the mass balance and  linear momentum balance (LMB) in the BFCS  $ \left\{\mathcal{E}_0\right\}$ are, respectively,
\begin{flalign}
\nabla\cdot\bm{u}=0 \hspace{0.5cm}
\text{and} \hspace{0.5cm}
    \rho\frac{\partial{\bm{u}}}{\partial t}+\nabla\cdot(\rho \bm{u}\otimes\bm{u})&=\nabla\cdot\bm{\sigma}+\rho(\bm{b}-2\bm{\omega}\times \bm{u}-\bm{r}\times\bm{\alpha}-\bm{A}_c),\label{eq:ns_1}
\end{flalign}
where $\bm{u}$ is the flow's velocity relative to $\mathcal{E}_0$, $\rho$ is the regolith's mean density, $\bm{\sigma}$ is the stress  tensor, ${\bm \omega}$ and $\bm{\alpha}$ are, respectively, the CB's angular velocity and angular acceleration, $\bm{r}$ is a mass element's  position vector relative to $C$,   $\bm{b}=\bm{b}_0-\bm{\omega}\times(\bm{\omega}\times \bm{r})$ is the total effective gravity in terms of  the CB's gravity $\bm{b}_0$ and the centrifugal acceleration $\bm{\omega}\times(\bm{\omega}\times \bm{r})$, $2\bm{\omega}\times \bm{u}$  is the Coriolis acceleration and $\bm{A}_C$ is the inertial acceleration of the CB's CoM. To complete \eqref{eq:ns_1} we need to prescribe a rheology appropriate for the granular continuum that relates the stress tensor to the flow's strain rate, and we  return to this in the next section.

Next, we consider how the CB's dynamics couples to the regolith flow. In the inertial CS $\mathcal{P}$, the LMB of the full system comprising of the CB and the flowing regolith  is 
\begin{equation}
    \frac{\td}{\td t}({M_c\bm{U}_{c}})+\frac{\td}{\td t}({M_r\bm{U}_{r}})=\bm{F}-\bm{F}_s, \label{eq:lmb_1} 
\end{equation}
where the subscripts `$c$' and `$r$' indicate the CB and the flowing regolith, respectively,  $M$ is the mass, $\bm{U}$ is  the CoM velocity, $\bm{F}$ represents external forces  on the system and $\bm{F}_s$ is the rate of momentum carried away by the mass shed from the body's surface. Due to erosion and deposition during landsliding, mass is exchanged between the stationary regolith (CB) and the flowing grains, so that  $M_c$ and $M_r$ may be modified. Additionally, $M_r$ may also change due to surface mass shedding.

Finally, in the inertial CS $\mathcal{P}$, the global angular momentum balance (AMB) of the system about the point $P$ is
\begin{equation*}
    \frac{\td \bm{H}_c}{\td t}+\frac{\td\bm{H}_r}{\td t}=\bm{T}- \bm{T}_s, 
\end{equation*}
where the subscripts `$c$' and `$r$' are as above, $\bm{H}$ represents angular momentum, $\bm{T}$ is the external torque and $\bm{T}_s$ is the rate at which  angular momentum is carried away by the shed mass. The above AMB may be simplified further 
% in the supplementary material 
to obtain the following form: 
\begin{equation}
\bm{\omega}\times(\bm{I}\cdot\bm{\omega}+\bm{h}_r)+\bm{I}\cdot\bm{\alpha}+\Dot{\bm{h}}_r+\Dot{\bm{I}}\cdot\bm{\omega}+ \bm{R}_{O} \times M\bm{A}_c ={\bm{T}}-{\bm{T}}_s,\label{eq:AMB}
\end{equation}
where $\bm{I}$ is the moment of inertia about $C$ of the system comprised, in general, of a static part (CB) and flowing regolith, $\bm{h}_r$ is the angular momentum of the flowing regolith about  $C$ in the BFCS $\mathcal{E}_0$,   $\Dot{(\ )}$ represents time differentiation in the BFCS,  $M$ and $\bm{R}_O$ are, respectively, the total system mass and the location of its CoM with respect to $C$. The first term on the left hand side of \eqref{eq:AMB} arises from the tumbling motion of the CB, the second is the rotational inertia, the third and fourth terms are due to  regolith flow and the final   
% In their work on Bennu, \cite{banik_top} employed only this term in their AMB, because they assumed that the body is always in pure spin about its axis of maximum moment of inertia. 
term is because the CoM of the system  may not be at the origin $C$, which introduces an extra contribution from its acceleration. 
 % In passing, we note that to obtain \eqref{eq:AMB} we took the inertial frame's origin and its velocity to momentarily coincide with that of the BFCS. 
%In  subsequent sections, we will assume nearly pure spin about a principal axis and the regolith layer to be much thinner than the body's extent. This will allow us to ignore terms in group  $\rm I$ and $\rm II$.

Thus far, while developing our modeling framework, we have made no assumptions regarding  forces and torques affecting our system, nor have we imposed any constraints on the kinematics of the CB or the nature of landslides. We may, in principle, implement this framework computationally. However, in the next section we introduce several physically motivated simplifications that make the presentation accessible, without sacrificing essential details or negating practical application. 

\vspace{-0.5cm}
\section{A simplified model}
% y the landslide model in two ways, viz. we take the landslides to be `thin', i.e. the flow's vertical depth is small compared to its lateral spread, and we assume that all landslides are axisymmetric and global. While the former restriction is natural, {the later is motivated  by the fact that the assumption of axisymmetry reduces one dimension in the problem and makes it easy to understand its dynamics. Also, many Near Earth Asteroids (NEAs)  are found to be top shaped and close to axisymmetry see Fig. } \ref{fig:top}). We aim to relax this in the future. 

Figure~\ref{fig:schematic} shows asteroids Bennu and Ryugu that were the targets of recent missions. Their shapes are observed to be fairly axisymmetric and the objects  were found to be in pure spin about the symmetry axis. This suggests specializing our model to an axisymmetric situation, wherein the initial body and all subsequent landslides are axisymmetric about a space-fixed rotation axis. We further assume that the landslides are shallow in that, their depth, normal to the body's topography, is much smaller than the extent of their spread over the body's surface. Consequently, the momentum carried by the regolith flow and shed mass is much smaller than the system's orbital momentum, which then balances the Sun's gravitation force on the right in \eqref{eq:lmb_1}.
% upon the displacement ${\bm R}_O$ of the system's CM from the CB's CoM is negligible and, so, we set  ${\bm R}_O = {\bm 0}$. 
Angular acceleration due to changes in the body's moment of inertia introduced by landsliding is, however, retained because, while  mass varies as  $D^2H$, the inertia does so in proportion to $D^4H$, where $D$ is the body's diameter and $H$ is typical depth of the landslide. Indeed, \cite{banik_top} found that, although the change in the body's rotation due to each landslide over its lifetime is small, the  cumulative effect is significant. 

 Now, the external forces $({\bm F})$ and moments $({\bm T})$ entering into the governing equations \eqref{eq:lmb_1} and \eqref{eq:AMB}  arise from  gravitational attraction of the Sun and other bodies, (thermal) radiation pressure and radiation torques. Radiation pressure  -- the Yarkovsky effect  --  influences  orbital motion, while radiation torques -- also called YORP torques -- change the body's rotation, especially over long time periods \citep{vokrouhlicky2015yarkovsky}. Here, we limit ourselves to small isolated objects. Thus, the Sun's quadrupole torque is negligible, as are the tidal interactions. Furthermore, the change in the body's orbit over its lifetime do not introduce significant variations in YORP torques. Along with our assumption above regarding the balance of orbital inertia and Sun's gravitational force in \eqref{eq:lmb_1}, these considerations decouple the angular momentum balance \eqref{eq:AMB} from  \eqref{eq:lmb_1}. Thus, we may study landslides independently of the body's orbital motion and external gravity forces. 

% External forces $({\bm F})$ and moments $({\bm T})$ entering into the governing equations \eqref{eq:lmb_1} and \eqref{eq:AMB}  arise primarily from  (thermal) radiation pressure, radiation torques and tidal interactions. We limit ourselves to isolated objects, so that  tidal interactions do not play a role. Radiation torques -- also called YORP torques \citep{vokrouhlicky2015yarkovsky} -- change the body's rotation, especially over long time periods, so that we retain ${\bm T}$ in \eqref{eq:AMB}. Finally,  radiation forces  -- the Yarkovsky effect \citep{vokrouhlicky2015yarkovsky} --  influence  orbital motion. However, at present, we assume no coupling between the body's rotational and orbital dynamics; hence  we  set ${\bm F} = {\bm 0}$ and also ignore the effect of mass shedding. Along with the assumption above that ${\bm A}_c = 0$, we then find that the global LMB \eqref{eq:lmb_1} is identically satisfied. We may thus study landslides independently of the body's rectilinear motion.

The AMB \eqref{eq:AMB} is further simplified for bodies that are nominally in a state of pure spin, i.e. bodies with a space-fixed rotation axis $\hat{\bm{e}}_3$,  which is generally their axis of greatest moment of inertia. This is true for many small planetary objects, including the asteroids in Fig.~\ref{fig:schematic}.  The reason for this may be traced back to the effect of internal friction that damps out tumbling motion in bodies in free space, driving them into a state of pure spin about their axis of greatest inertia \citep{sharma2005nutational}. 

For a body in pure spin about $\hat{\bm e}_3$ we have $\bm{\omega}=\omega\hat{\bm{e}}_3 $ and $\bm{\alpha}=\alpha\hat{\bm{e}}_3.$ Invoking this and the  simplifications above, we arrive at the following coupled equations for shallow, incompressible regolith flow on a planetary body in pure spin:
\begin{flalign}
&& &&\nabla\cdot \bm{u}&=0,\label{eq:cont_2} &&\\
&& && \rho\Dot{\bm{u}}+\nabla\cdot(\rho \bm{u}\otimes\bm{u})&=\nabla\cdot\bm{\sigma}+\rho(\bm{b}-2\bm{\omega}\times \bm{u}-\bm{\alpha}\times\bm{r})\label{eq:ns_2} &&\\
&&\text{and} &&I\alpha + \dot{h}_r + \dot{{I}}{\omega}&= T - {T}_s \label{eq:amb_3}, &&
\end{flalign}
 where $I$ and $h_r$ are, respectively, the system's moment of inertia and the flowing regolith's angular momentum  about the rotation axis $\hat{\bm e}_3$, $\Dot{{I}}_s$ and ${\dot{h}}_s$  are the rate of change of the moment of inertia and angular momentum due to mass shedding and $T-{T}_s = \left({\bm T} - {\bm T}_s\right)\cdot\hat{\bm e}_3$. 
 % We emphasize that all quantities and their rates of change in the above equations are measured in the body's BFCS.

We also require  a rheology of the flowing regolith and boundary conditions (BCs) to close the set \eqref{eq:cont_2} and \eqref{eq:ns_2}.  We introduce the former   in Sec.~\ref{sec:da} below, but discuss the latter here. We assume no erosion / deposition of material onto / from the CB during a landslide. At the free top surface $F^s(\bm{r})=0$, identified by the superscript $s$, we impose the vanishing of the traction and the usual kinematic conditions, respectively,
\begin{equation}
  \boldsymbol{\sigma}^s\cdot\boldsymbol{n}^s=0\hspace{0.5cm}\text{and}\hspace{0.5cm}  \frac{\partial F^s}{\partial t} +\left(\bm{u}\cdot\nabla\right) F^s = 0,\label{eq:bc_1}
\end{equation}
where $\boldsymbol{n}^s$ is the normal to the free surface. Mass shedding is assumed to commence when the pressure at the flow's base, denoted by the superscript $b$, becomes tensile in nature, i.e. when $\boldsymbol{n}^b\cdot\boldsymbol{\sigma}^b\cdot \boldsymbol{n}^b>0$.   Now, the flow adheres to the basal topography and the surface traction there is assumed to be governed by Coulomb's friction law. The mathematical statements of these boundary conditions at the flow's base are, respectively, 
\begin{equation}
\boldsymbol{u}^b\cdot\boldsymbol{n}^b=0 \hspace{0.5cm} \text{and}\hspace{0.5cm}
    % \boldsymbol{\sigma}^b\cdot\boldsymbol{n}^b=(\boldsymbol{n}^b\cdot \boldsymbol{\sigma}^b \cdot \boldsymbol{n}^b)\left( -\mu\frac{\boldsymbol{u}^b}{|\boldsymbol{u}^b|} +\boldsymbol{n}^b\right)
    \left({\bm 1} - \boldsymbol{n}^b\otimes\boldsymbol{n}^b\right)\cdot\boldsymbol{\sigma}^b\cdot\boldsymbol{n}^b =-\mu\boldsymbol{n}^b\cdot \boldsymbol{\sigma}^b \cdot \boldsymbol{n}^b\frac{\boldsymbol{u}^b}{|\boldsymbol{u}^b|}, \label{eq:bc_2}
\end{equation}
where $\mu$ is the basal friction coefficient. Landslides flow over the CB, which is comprised of the same type of grains, so that we set $\tan^{-1}\mu = \zeta$, the regolith's internal friction angle. 

Now, limiting ourselves to  shallow axisymmetric landslides on an initially axisymmetric body allows us to reduce the spatial dimensionality of the governing equations to {\em one} -- derivatives in the longitudinal direction vanish and we may depth-average our equations to suppress through-thickness variations in the regolith flow. Furthermore, because the regolith flow's thickness $H$ is much smaller compared to the CB's mean radius $R_{CB}$, we may now introduce the small parameter $\varepsilon := {H}/{R_{CB}}\ll 1$. This allows us to approximate various quantities by ignoring variations that are $\mathcal{O}(\varepsilon)$ or smaller.  These steps simplify our mathematical description considerably, as we now see. 

 % \eqref{eq:cont_2}-\eqref{eq:amb_3} and BCs \eqref{eq:bc_1}-\eqref{eq:bc_2}, 

% \subsection{Local axisymmetric coordinate system}

\subsection{Depth-averaging}\label{sec:da}

The regolith flow's shallowness prompts us to integrate the LMB \eqref{eq:ns_2} through the flow's thickness \citep{Luca2016, gray1999gravity}. To facilitate this we  introduce the natural  axisymmetric coordinate system associated with the body's shape in which to express our governing equations; see, for example, the schematic in Fig.~\ref{fig:schematic}c. The CB's axisymmetric surface is defined by its distance $r_{CB}=R(\theta)$ from $C$ in terms of the co-latitude $\theta$. To describe the regolith flow over the CB's surface, we introduce the  contravariant coordinate system $(x^1, x^2, x^3) = (n,\theta,\phi)$, where $n$ is the normal distance from the CB's surface and  $\phi$ is the  longitude. A material point {\em in} the regolith flow is then located by 
\begin{equation}
    \bm{r} = r_{CB}\hat{\bm{e}}_r +n\hat{\bm n}=R(\theta)\hat{\bm{e}}_r +n\hat{\bm n}, \hspace{1cm} \text{where}\hspace{1cm} \hat{\bm{ n}} = N^{-1}{\left(R\hat{\bm{e}}_r - R'\hat{\bm{e}}_\theta\right)}, \label{eq:position}
\end{equation}
  is the outward unit surface normal, $N=\sqrt{R^2 + R'^2}$, $\hat{\bm{e}}_i$  are the unit vectors in spherical coordinates and  prime $(')$ denotes differentiation with respect to $\theta$. Finally, the   natural basis vectors $\bm{g}_i$ are obtained as
$
    \bm{g}_i={\partial \bm{r} }/{\partial x^i} \label{eq:curv_1}
$:
\begin{flalign}
    &\bm{g}_1 = \hat{\bm{ n}}, \hspace{0.25cm} 
    \bm{g}_2  = \left(1+{n}/{\mathcal{R}_\theta}\right)(R'\hat{\bm{e}}_r+R\hat{\bm{e}}_\theta)\hspace{0.25cm} 
    \text{and} \hspace{0.25cm}  \bm{g}_3  = \mathcal{R}_\phi(1+\kappa_n n)\hat{\bm{e}}_\phi, \hspace{0.5cm} \label{eq:basis}\\
  \text{where} &\hspace{1cm}  \mathcal{R}_{\theta } ={N^{3}}\left({R^2+2R'^2-RR''}\right)^{-1} \hspace{0.5cm}\text{and} \hspace{0.5cm} \mathcal{R}_\phi = {R\sin\theta}&&
\end{flalign}
    are, respectively, radii of curvature of $\theta-$ and $\phi-$ coordinate curves on the CB's  surface and  
$\kappa_{ n}$ and  $\kappa_{ g}$ are the normal and geodesic curvatures of $\phi-$curves on the CB's surface, respectively, which are given by
\begin{equation}
\kappa_{ n} = -{(R\cos{\theta})'}/{J}\hspace{0.5cm}\text{and} \hspace{0.5cm}\kappa_{ g} = {\mathcal{R}_\phi'}/{ J},\hspace{0.5cm}
\end{equation}
where $J=N \mathcal{R}_\phi$ is the Jacobian of the transformation from Cartesian to axisymmetric coordinates at the CB's surface.

 Before proceeding we non-dimensionalize our equations as per the following definitions:
 \begin{equation}
\begin{array}{cc}  \left\{n,t\right\}=\left\{H\hat{n},\sqrt{\displaystyle{R_{CB}}/{b}}\,\hat{t},\right\}; \hspace{0.5cm}
\left\{u_n,u_\theta,u_\phi\right\}=\sqrt{bR_{{CB}}}\left\{\varepsilon\hat{u}_n,\hat{u}_\theta,\hat{u}_\phi\right\};\\ 
\left\{\omega,T\right\}=\left\{\sqrt{b/R_{CB}}\,\hat{\omega},\rho b {R^4_{CB}}\,  T\right\}; \hspace{2mm}
\text{and} \hspace{2mm} \left\{\sigma^d_{ij},\sigma^o_{ij}\right\}=\rho b H\left\{\hat{\sigma}^d_{ij},\mu\hat{\sigma}^o_{ij}\right\},\label{eq:non_dim}
\end{array}
\end{equation}
where $\sigma^d_{ij}$ and $\sigma^o_{ij}$ are the diagonal and off-diagonal components of ${\bm \sigma}$, $b$ is the gravitational acceleration on the surface of a sphere of radius $R_{CB}$ and $\hat{(.)}$ identifies non-dimensional quantities; the $\,\hat{}\,$ is subsequently dropped for simplicity.

We now define the depth-averaged quantities, identified by $\bar{(\cdot)}$, 
\begin{equation}
   \left\{\overline{v}(\theta,t), \overline{w}(\theta,t)\right\} = \frac{1}{h}\int^{h}_{0}\left\{ v(n,\theta,t),  w(n,\theta,t)\right\} \, \td n \ \ 
   \text{and} \ \ 
   \overline{vw} = \beta_{vw}(\theta,t) \overline{v} \,\overline{w}, \label{eq:vw}
\end{equation}
where $h$ is the flow's depth and the  depth-variation parameter $\beta_{vw}$ is introduced to express the average of the product of velocities as the product of their individual averages. A plug-like velocity profile is a fairly good assumption for granular flows \citep{gdr2004dense}. Consequently, we approximate $\beta_{vw}$ as unity, as did  \cite{gray1999gravity}. This allows us to equate  the depth-averaged velocity to the basal velocity, i.e. $\bar{\bm u}= {\bm u}^b$. 
 % \item \textit{Rheology:} 
 
 To close our mathematical description we  need a rheological description of the shallow regolith flow. We express the stress tensor as $\bm{\sigma} =-p\bm{1} +\bm{\tau}$, where $p$ is the pressure and ${\bm \tau}$ is the deviatoric stress tensor.
 \cite{gray2014depth} and \cite{baker2016two} proposed a depth-averaged constitutive law of the form
\begin{equation}
    h\bar{\bm{\tau}} =\rho \nu h^{3/2} \bar{\bm{D}},
\end{equation}
where $\nu$ and $\bar{\bm{D}}$ are the depth-averaged viscosity and strain rate tensor, respectively. 
% When replaced in the depth-averaged LMB, the  term $\bar{\bm \tau}$ acts as a diffusive viscous term.
However, the contribution from $\bar{\bm \tau}$ in the depth-averaged LMB is generally rather small \citep{baker2016two,rocha2019self}. Given this, for simplicity, here we ignore $\bar{\bm \tau}$ and equate $\overline{\sigma}_{\theta\theta}=\overline{\sigma}_{\phi\phi} =\overline{\sigma}_{nn}=-\overline{p}$.
% Other constitutive laws may be found in \cite{Luca2016constitutive}.

The depth-averaged momentum equation in the $n-$direction  now becomes
\begin{flalign}
    &&p &= \psi (h-n) +O(\varepsilon), 
    \label{eq:psi}&& \\
    \text{where} &&
   \psi = -(f_n +{\overline{u}_\theta^2}{\mathcal{R}_\theta}^{-1} + \kappa_{ n}\overline{u}^2_{\phi}) \ \ 
   &\text{and} \ \ 
   f_n=b_n+\mathcal{R}_\phi{\kappa_{ n}}\left(\mathcal{R}_\phi{\omega^2} +2\omega \overline{u}_\phi\right) && \label{eq:psidef}
\end{flalign}
is the body force term. 
As mentioned, mass shedding commences when basal pressure becomes negative which, from \eqref{eq:psi}, is equivalent to the requirement that  ${\psi}<0$.

Incorporating the above, our final depth-averaged  equations governing shallow regolith flow  when mass shedding is absent, i.e. ${\psi}>0$, become
\begin{flalign}
    &\frac{\partial }{\partial t}( J h) +  \frac{\partial }{\partial \theta}\left(\mathcal{R}_\phi{\overline{u}_\theta} h\right) =0,\label{eq:fin_cont}&&\\
    &\frac{\partial }{\partial t}( J \overline{u}_\theta h) + \frac{\partial}{\partial \theta}\left\{\left({\overline{u}_\theta^2}  +\frac{\varepsilon}{2}h\right)\mathcal{R}_\phi h\right\} =\Bigg[-\frac{\mu \psi\overline{u}_\theta}{\sqrt{\overline{u}^2_\theta + \overline{u}^2_\phi}} + b_\theta  ~ . . . \nonumber \\
    & \hspace{3.5cm}...  +~  \kappa_{ g}\left\{\mathcal{R}_\phi{\omega} \left(\mathcal{R}_\phi{\omega} +2\overline{u}_\phi\right)+\overline{u}^2_\phi + \varepsilon{\psi}\frac{h}{2}\right\} \Bigg]J h ,\label{eq:fin_mom1}&&\\
    \text{and}\hspace{0.20cm}  &\frac{\partial }{\partial t}\left( {J \mathcal{R}_\phi \overline{u}_\phi} h \right) + \frac{\partial}{\partial \theta}\left(\mathcal{R}^2_\phi{\overline{u}_\theta \overline{u}_\phi} h\right) = -\left[\frac{\mathcal{R}_\phi\mu {\psi}{\overline{u}}_\phi}{\sqrt{\overline{u}^2_\theta + \overline{u}^2_\phi}}  +  \mathcal{R}^2_\phi\left(2{\kappa_{ g}}\omega \overline{u}_\theta +{\alpha}\right)\right]{J} h,\label{eq:fin_mom2}&&
\end{flalign}
where  the Jacobian, curvatures and body forces  are evaluated at the CB's surface, which introduces  errors only at $O(\varepsilon).$ The equations above have to be solved for the flow rate $\left\{{\bar u}_\theta, {\bar u}_\phi\right\}$ and depth $h$ of the landslide. Finally,  mass is shed from regions where   ${\psi}<0$. In those regions  we set 
 \[ h=\overline{u}_\theta=\overline{u}_\phi=0.\]

\subsection{Rotational dynamics}
It remains to follow the evolution of the body's rotation rate $\omega$. The angular momentum balance equation \eqref{eq:amb_3} after non-dimensionalization and invoking axisymmetry becomes 

\begin{flalign}
     &\alpha = \frac{1}{I}\left\{T-{2\pi\varepsilon}\int_{\psi\geq0} {J}\mathcal{R}_\phi\frac{\partial}{\partial t}\left(h{\overline{U}}_\phi\right) \,\td\theta\right\} +O(\varepsilon^2),&&\label{eq:AMB_1}\\
      \text{where}\hspace{1cm}  & I = {2\pi}\int^\pi_0\left(\mathcal{R}_\phi^3R^2/5+\varepsilon\mathcal{R}_\phi^2{Jh}\right)\,\td \theta \hspace{1cm} \text{and}\hspace{1cm} \overline{U}_\phi = \overline{u}_\phi +\mathcal{R}_\phi{\omega}&&\nonumber
 \end{flalign}
  are the system's moment of inertia and the regolith's  azimuthal velocity in the inertial frame, respectively. The integral in \eqref{eq:AMB_1} is over those regions where the basal pressure does not vanish.  From \eqref{eq:AMB_1} we may determine the angular acceleration $\alpha$ at any time in terms of the average azimuthal velocity $\bar{u}_\phi$ and the depth $h$ of the landslide. Time integration of $\alpha$ yields $\omega$.

\section{Results and discussion}
\begin{figure}
    \centering
    \includegraphics[width=0.7\linewidth]{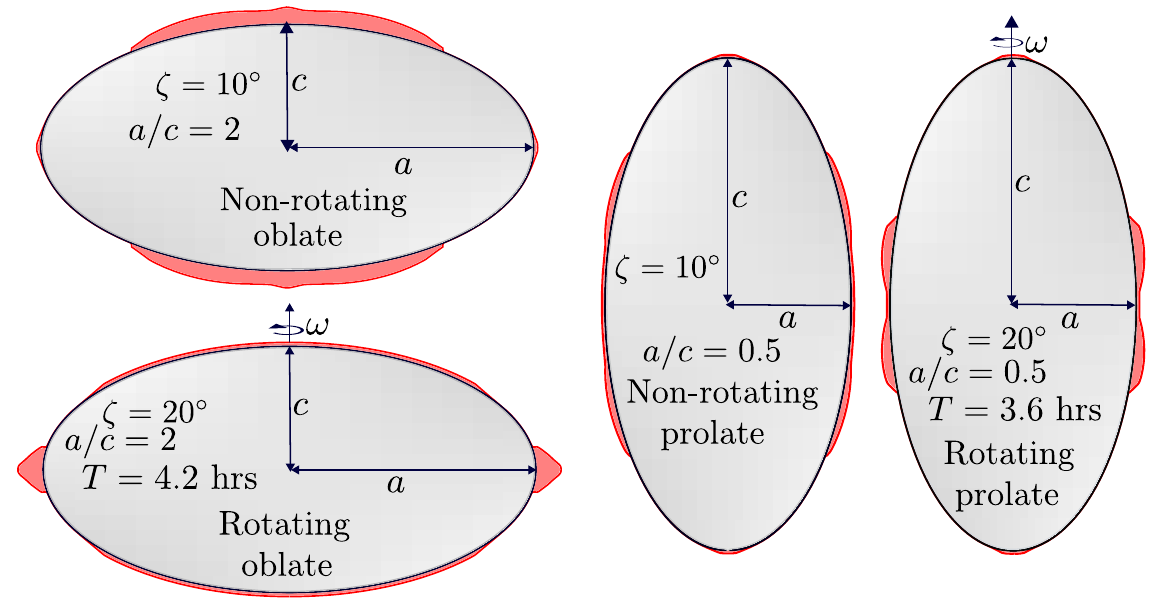}
    \caption{Final regolith deposited by landsliding of an initially uniform regolith layer on stationary and rotating oblate and prolate ellipsoids. The regolith height is magnified two times for clarity.}
    \label{fig:oblate}
\end{figure}
To begin with, we simulate regolith flows on bodies in the limit of the body's rotation being zero and very fast to understand, respectively, gravity- and rotation- driven flows.  We initiate the flow by distributing   regolith  uniformly  on the body's surface with depth $H=\varepsilon R_{CB}$, where we  set $\varepsilon = 0.01$.  The flow of the ensuing landslide is found by numerically solving  \eqref{eq:fin_cont}-\eqref{eq:fin_mom2} utilizing the non-oscillatory central schemes of \cite{kurganov}. The gravity field of the axisymmetric body is found by slicing the body into  infinitesimal discs along the symmetry axis and then adding  their individual contributions \citep{hure2012key}.

Figure~\ref{fig:oblate} presents the final profile of initially uniformly distributed regolith layer on an oblate and a prolate body that were either stationary or rotating. When rotation is absent, the gravitational field is too weak to overcome internal friction; thus results are shown for a small friction angle $\zeta=10^\circ$. We find that landslides transport grains  to the poles on an oblate body and to the equator for a prolate body. In contrast,  rotational effects always drive grains  to the equator. 
% Oblateness of the body influences the extent of this motion. For an oblate body the migration is limited near the equator while for a prolate body it is more near poles. 
Thus, for a prolate body, rotation assists the gravity-driven motion augmenting grain accumulation at the equator. On an oblate body, rotation and gravity compete, and we may find comparable amount of regolith at both poles and the equator. We remark that rotation about the long axis of a prolate body is generally unstable. Thus, that example may be treated as illustrative. 

\begin{figure}
    \centering
    \includegraphics[width=0.8\linewidth]{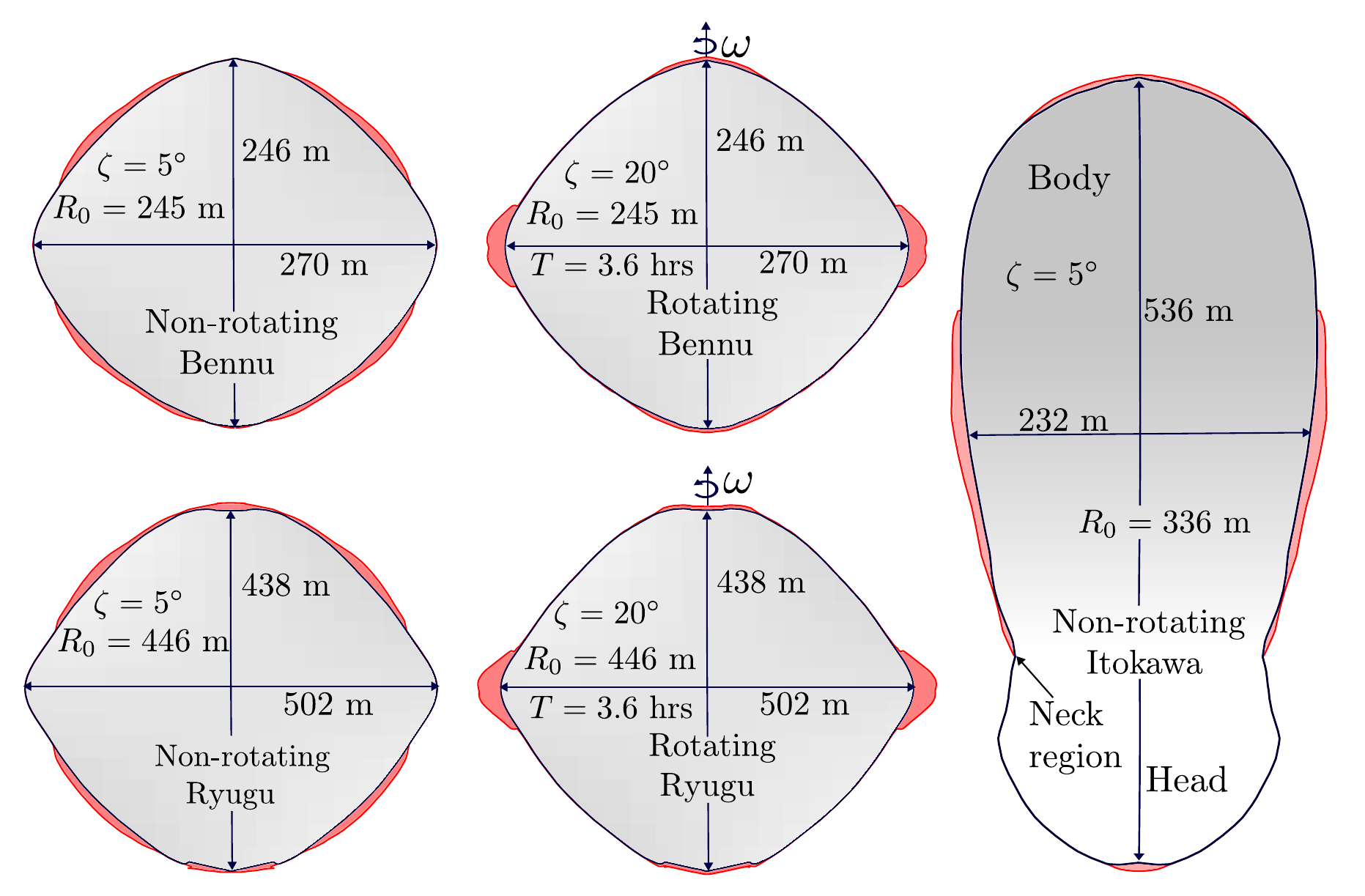}
    \caption{ Final regolith deposited by landsliding of an initially uniform regolith layer  on axisymmetrically modeled asteroids Bennu, Ryugu and Itokawa. The regolith thickness is magnified two times for clarity. }
    \label{fig:bennu}
\end{figure}

We consider now applications to small planetary objects. For this we create axisymmetric bodies  from shape models of actual nominally axisymmetric asteroids  by averaging their surface profile across longitudes at a fixed latitude. Figure~\ref{fig:bennu} shows final deposits of initially uniformly spread regolith on  asteroids  Bennu and Ryugu from Fig. \ref{fig:schematic} and Itokawa. Were Bennu and Ryugu not rotating, then we would find  grains accumulated at mid-latitudes; this is where the  equilibria at the surface of such top-shaped bodies are located \citep{banik_top}. Interestingly, we find grains deposited around the poles of Ryugu due to presence of a local topographical depression. At a higher rotation rate, we find that regolith accumulates at the equator, as in the examples in Fig.~\ref{fig:oblate}. Finally, while Itokawa may be somewhat crudely approximated as an axisymmetric body, it is observed to rotate very slowly with a time period of 12 hours about its short axis. Thus, to fit into our modeling we here investigate landsliding on a {\em non}-rotating Itokawa. We find that, while regolith migrates towards Itokawa's poles,  a significant fraction is captured in the depression around the ``neck" region where Itokawa's ``head" and ``body" are joined. 

Finally, we explore one instance of the evolution of the shape and spin state of an initially spherical rubble asteroid over its lifetime due to impact-induced landsliding and the action of YORP torque. We assume that the asteroid experiences 143 landslide events, which is consistent with the statistically expected number of impact events that are large enough to excite global landslides \citep{ghosh2024segregation}. Following \cite{banik_top} we assume that each landslide had a thickness ratio $\varepsilon=0.01$.  The YORP torque equalled that experienced by the  Bennu were it  located in the main belt \citep{bottke2015search}. 

\begin{figure}
    \centering
    \includegraphics[width=0.8\linewidth]{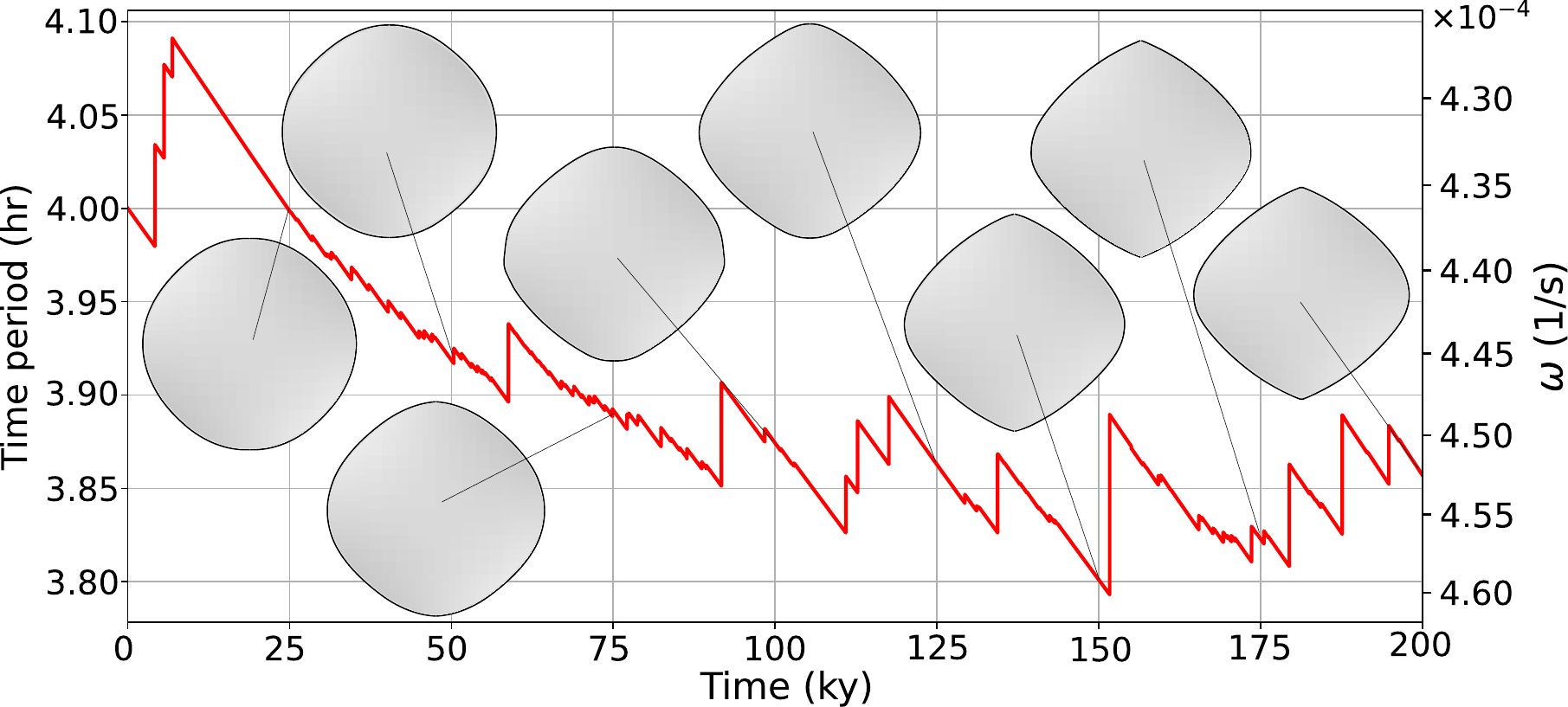}
    \caption{The  shape and spin  evolution of an initially spherical rubble asteroid due to YORP torque and impact-induced landslides over 200  kilo-years (ky). The basal and internal friction angle $\zeta=20^\circ$.}
    \label{fig:evolution}
\end{figure}

Figure \ref{fig:evolution} reports how over its lifetime the asteroid's rotation rate changes and its shape evolves into a top shape, similar to the asteroids  in Fig.~\ref{fig:schematic}. The body's shape is progressively changed due to deposition from multiple landslide events, as indicated in Fig.~\ref{fig:schematic}d. The spin is raised continuously by the YORP torque, while each landsliding event lowers the spin due to redistribution of regolith to the equator. This stops the spin from increasing beyond a limit and prevents rotational fission. This is an important finding, as it suggests that rotational fission may not be as prevalent, if at all, in the evolution of small rubble planetary bodies as presently supposed.  Furthermore, the formation of top shapes within a time-scale of $100$ kiloyears (ky) may explain the abundance of top-shaped asteroids in near-Earth orbits. 

\vspace{-0.5cm}
\section{Conclusion}
We have presented a framework for studying regolith motion on axisymmetric bodies incorporating effects of spin change, non-uniform gravity, varying topography and mass shedding employing depth-averaging. These are then coupled to the body's rotational dynamics.  
% The angular momentum equation shows that the change in spin is due to frictional interaction between grains and central body and is independent of mass shedding. 
These equations are then solved to investigate regolith motion on axisymmetric shapes derived from real asteroids Bennu, Ryugu and Itokawa. We also demonstrated  our framework's unique ability in following shape and spin evolution of such bodies over long time scales relevant to planetary bodies. We find that shape changes due to landsliding may suppress rotational fission, contradicting current understanding.   Furthermore, top-shapes may form rapidly and this may explain the overabundance of top-shaped NEAs. We are presently  confirming these  results by running a statistically large number of simulations. 

There are two immediate avenues of development.  First,  DE simulations of \cite{song2024integrated} report that  shed mass re-impacts the surface. However, we assume  that shed mass is lost to infinity. Consequently, we over-predict mass loss; indeed $17.5\%$ of total mass in the computation  in Fig. \ref{fig:evolution} was lost. Thus, mass shedding has to be modeled better. Second, we need to relax our assumption of axisymmetry. This will allow us to address both small-scale impact events that initiate local landslides, as well as bodies of more general shapes. 

Looking ahead, our framework may be modified to analyze specific future events: the effects of Earth's tidal forces on asteroid Apophis during its close encounter in 2029, and the surface changes on asteroid Didymos caused in 2022 by the Double Asteroid Redirection  Test (DART) mission, to study which the Hera space mission was launched in 2024.

\begin{bmhead}[Acknowledgements.] 
KG thanks financial support of PMRF, India. IS acknowledges MATRICS grant SERB/F/11747/2023-2024.
\end{bmhead}
\begin{bmhead}[Declaration of interests.]
The authors report no conflict of interest.
\end{bmhead}
\bibliographystyle{jfm}
\bibliography{bibliography}

\appendix

\end{document}